\documentclass[pra,twocolumn,showpacs,preprintnumbers,amsmath,amssymb]{revtex4}

\usepackage{graphicx}
\usepackage{bm}

\newcommand{\ket}[2][]{{|#2\rangle_{#1}}}
\newcommand{\bra}[2][]{{}_{#1}\langle #2|}

\newcommand{\proj}[2][]{\ket{#2}_{#1}\bra{#2}}
\newcommand{\Tr}{\textrm{Tr}}

\def\>{\rangle}
\def\<{\langle}
\def\ot{\otimes}

\def\duzomniejsze{<\kern-.7mm<}
\def\duzowieksze{>\kern-.7mm>}

\def\textbf#1{{\bf #1}}
\def\beq{\begin{equation}}
\def\eeq{\end{equation}}
\def\be{\begin{equation}}
\def\ee{\end{equation}}
\def\ben{\begin{eqnarray}}
\def\een{\end{eqnarray}}
\def\beqa{\begin{eqnarray}}
\def\eeqa{\end{eqnarray}}
\def\eea{\end{array}}
\def\bea{\begin{array}}
\newcommand{\bei}{\begin{itemize}}
\newcommand{\eei}{\end{itemize}}
\newcommand{\bee}{\begin{enumerate}}
\newcommand{\eee}{\end{enumerate}}

\begin{document}


\title{Quantum privacy witness}

\author{Konrad Banaszek}

\affiliation{Instytut Fizyki Teoretycznej, Wydzia\l{} Fizyki, Uniwersytet Warszawski, Ho\.{z}a 69, PL-00-681 Warszawa, Poland}

\author{Karol Horodecki }

\affiliation{Instytut Informatyki, Universytet Gda\'{n}ski, 80-952 Gda\'{n}sk, Poland}

\author{Pawe\l{} Horodecki}
\affiliation{Wydzia\l{} Fizyki Technicznej i Matematyki Stosowanej, Politechnika Gda\'{n}ska, 80-952 Gda\'{n}sk, Poland}

\date{\today}

\begin{abstract}
While it is usually known that the mean value of a single observable is enough to detect entanglement
or its distillability, the counterpart of such an approach in the case of quatum privacy has been missing.
Here we develop the concept of a privacy witness, i.e.\ a single observable that may detect presence
of the secure key even in the case of bound entanglement. Then we develop the notion
of secret key estimation based on few observables and discuss the witness decomposition
into local measurements. The surprising property of the witness is that
 with the help of a low number of product mesurements involved
it may still report the key values that
are {\it strictly above} distillable entanglement of the state. For an exemplary four-qubit state studied in a recent experiment
[K. Dobek {\em et al.}, Phys. Rev. Lett. {\bf 106}, 030501 (2011)] this means $6$ Pauli operator product measurements versus $81$ needed to carry out the complete quantum state tomography. The present approach may be viewed as a paradigm for
the general program of experimentally friendly detection and estimation
of task-dedicated quantum entanglement.

 \end{abstract}

\maketitle

\section{Introduction}
Entanglement based cryptography \cite{Ekert91}, equivalent formally to the BB84 scheme \cite{BB84}, is
the one that uses the power of quantum entanglement monogamy obeyed by
a maximally entangled pure quantum state. If the state is noisy then in some cases
it is possible to run an entanglement distillation process \cite{BBPS1996}
which may be interpreted as quantum privacy amplification \cite{QPA}.
Since the final output is maximally entangled, it may be used directly for secret key generation. The efficiency
of this procedure is quantified with distillable entanglement $E_D$, which defines how many singlet states can be obtained in the asymptotic
regime per one input.

Still it was known that certain states which cannot be prepared by local operations and classical communication (LOCC) are not distillable, exhibiting
the phenomenon of bound entanglement \cite{bound}. For a long time bound entanglement was believed to be useless for cryptography,
but several years ago it was shown \cite{pptkey,keyhuge},  that at least some bound entangled states may be useful in  quantum
cryptography. This is one extreme instance of the general fact that the amount of distillable secure key $K_{D}$ may exceed the
amount of distillable singlets $E_{D}$. The latter effect has been verified in a recent experiment \cite{pbits-exp}.

The key ingredient in the complete theory of distilling a secret key from quantum states
\cite{pptkey,keyhuge} is the notion of a {\it private bit}, p-bit --- or more generally a private dit, p-dit ---
which is a delocalized maximally entangled state that still retains some entanglement monogamy result.
A quantum p-dit is composed from a $d \otimes d$ key part $AB$, and the shield part $A'B'$, shared between Alice (subsystems $AA'$)
and Bob (subsystems $BB'$) in such a way that the local von Neumann measurements on the key part in a {\em particular}
basis will make its results completely statistically uncorrelated from the results of any measurement
of an eavesdropper Eve on her subsystem $E$, which is a part of the purification $|\Psi\rangle_{ABA'B'E}$ of the p-dit state $\varrho_{ABA'B'}$.
There is a nice explanation how the shield part protects the statistics of the measurement
on $A$ and $B$ to be correlated to Eve: it just makes it impossible to distinguish the
results of the measurement by an external observer.

An obvious way to determine privacy properties is to reconstruct tomographically the complete p-dit state $\varrho_{ABA'B'}$. This however is a very time consuming process, especially if the system under investigation is high-dimensional. The aim of the present paper is to give bounds on the distillable
secure key based just on few observables. This advances further the study presented in Ref.~\cite{smallkey}, where it was proposed to carry out a tomography of the so-called privacy-squeezed state of the state of merit. We demonstrate that
a single observable suffices to provide a nontrivial bound. We also provide more accurate estimates based on two observables.
These results provide tools for application-specific detection of entanglement, refining the fundamental concept of the
entanglement witness proposed in \cite{sep1996} and \cite{TerhalBell2000} that can be also subjected to optimization with respect to local
measurements \cite{GuehneLocalMeasurementWitness,GuehneHBE_s} and used to quantify the amount of entanglement \cite{Jaynes,Brandao2005-witent}.

The present results can be viewed as an outcome of a more
general research program: {\it experimentally friendly detection/estimation of task-dedicated
quantum entanglement and/or correlations}.
In fact it is quite usual that we are interested in that aspect of entanglement  which is useful for
specific quantum information task. The quantity characterizing this aspect may be a monotone but
we believe that it need not to be in general. For instance  it
is known that there are cases when specific Bell
inequalities which are important for device independent
cryptography are better violated by nonmaximally entangled states.
In this context we believe that the present paradigm will lead to
systematic development of  experimentally friendly detection/estimation of
resources for quantum information tasks.

This paper is organized as follows. In Sec.~\ref{sec:Keybounds} we elaborate on lower bounds on distillable entanglement and distillable key. In Sec.~\ref{sec:single} we present a lower bound on distillable key in terms of single parameter, i.e.\ single privacy witness. An approximate version of this bound is presented in Appendix. In Sec.~\ref{ref:double} we discuss how to infer privacy of a noisy state from the expectation values of two observables. Finally, Sec.~\ref{Sec:Conclusions} concludes the paper.

\section{Key bounds}
\label{sec:Keybounds}
Let us start by reviewing how an individual observable can be used to estimate distillable entanglement $E_D$.
The most natural observable in this context is a projector
\begin{equation}
\hat{W}_{\text{ent}} = \proj{\Psi_{\text{max}}}
\end{equation}
onto a maximally entangled state  $|\Psi_{\text{max}}\rangle =\frac{1}{\sqrt{d}} \sum_{i=1}^{d} |\phi_i\rangle_{A} \otimes |\psi_i \rangle_{B} $ of two $d$-level systems, where
$\{ \ket[A]{\phi_{i}} \}$  and $\{ \ket[B]{\psi_{i}} \}$ are any two orthonormal bases. Following the idea of Ref.~\cite{BBPS1996}
dealing with the case $d=2$, there is a protocol for an arbitrary $d$ such that if $F(\hat{\varrho}_{AB})=\Tr(\hat{W}_{\text{ent}} \hat{\varrho}_{AB})
=\langle \Psi_{\text{max}}|\hat{\varrho}_{AB}|\Psi_{\text{max}}\rangle$ satisfies $F(\hat{\varrho}_{AB}) > \frac{1}{d}$ then the state
$\hat{\varrho}_{AB}$ is distillable \cite{reduction}.

The corresponding rate of the distillation protocol can be easily estimated from below
by the hashing protocol \cite{DevetakWinter-hash-prl,DevetakWinter-hash} which gives
an estimate for the distillable entanglement as
\begin{equation}
E_{D}(\hat\varrho_{AB}) \geq S(\hat\varrho_{B}) - S(\hat\varrho_{AB})
\label{hashing}
\end{equation}
where $S(\cdot)$ denotes the von Neumann entropy.

Since an application of the  so-called $U \otimes U^{*}$ twirling \cite{reduction} can only decrease distillable entanglement $E_{D}$
we may twirl the state $\hat\varrho_{AB}$ in order to bring it to a highly symmetric form and then apply the hashing inequality (\ref{hashing})
which eventually gives:
\begin{equation}
E_{D}(\hat\varrho_{AB}) \geq \log_2 d - H\left(F,\frac{1-F}{d^{2}-1},..., \frac{1-F}{d^{2}-1}\right) ,
\label{fidelity}
\end{equation}
where $F = F(\hat\varrho_{AB})$ and $H(\{ p_i \} ) = -\sum_{i} p_i \log_2 p_i $ is the Shannon entropy.

The above formula is valid for any bipartite quantum state $\hat\varrho_{AB}$.
There are more sophisticated twirling protocols.
For instance for two qubits there is  a protocol \cite{BBPSSW1996} utilizing selected random Pauli operations that brings
the state to a form diagonal in the Bell basis:
\begin{equation}
\hat{\varrho}_{AB}^{\text{Bell}} = \sum_{i=1}^{4} p_i  \proj{\Psi_i}
\end{equation}
where $\ket{\Psi_{i}}$ are Bell states.
Applying the hashing protocol to $\hat{\varrho}_{AB}^{\text{Bell}}$ leads to:
\begin{equation}
\label{Eq:ED>=Bell}
E_{D}(\hat\varrho_{AB})\geq 1 - H(p_1,p_2,p_3,p_4).
\end{equation}
Eqs.~(\ref{fidelity}) and (\ref{Eq:ED>=Bell}) provide bounds on the key rate for $\hat\varrho_{AB}$, as distilled singlet states can be used for the standard Ekert protocol. In general however, this may be not the most efficient way to generate the key.

As pointed out in the introduction, a straighforward way to verify
that we have a p-bit or a state close to p-dit enough
that the secret key may be distilled. It is based on the so-called quantum state tomography which allows
to calculate specific entropic functions that can be used to estimate the amount of the secret key
from below. A useful quantity is the Devetak-Winter  function $K^{\rightarrow}_{DW}$ which quantifies
the secret key distillable through a specific one-way secret key distillation protocol:
\begin{equation}
K_{D} (\hat\varrho_{ABA'B'}) \geq K^{\rightarrow}_{DW} (\hat\varrho_{ABA'B'}).
\label{Devetak-Winter}
\end{equation}
The Devetak-Winter function is given explicitly by the difference of two Holevo quantities $\chi_{AB}$ and $\chi_{AE}$
which characterize the amount of information that can be recovered respectively by Bob and Eve from subsystems $B$ and $E$
after Alice carried out a von Neumann measurement of the subsystem $A$ in the secure basis $\{ \ket[A]{i} \}$:
\be
K^{\rightarrow}_{DW} (\hat\varrho_{ABA'B'})= \chi_{AB}\left(\Tr_{E}\hat\varrho_{ABE}^{\text{cqq}}\right)
- \chi_{AE} \left(\Tr_{B}\hat\varrho_{ABE}^{\text{cqq}}\right) .
\ee
The state $\hat\varrho_{ABE}^{\text{cqq}}$ is given by:
\begin{multline}
\hat\varrho_{ABE}^{\text{cqq}} = \Tr_{A'B'} \left( \sum_{i} \left( \proj[A]{i} \otimes \hat\openone_{A'BB'E} \right) \right. \\
 \left. \times \proj[AA'BB'E]{\Psi} \left( \proj[A]{i} \otimes \hat\openone_{A'BB'E} \right) \vphantom{\sum_{i}} \right).
\end{multline}

Let us now restrict our attention to a situation when the key part is composed of two qubits $A$ and $B$. The complete density matrix
$\hat\varrho_{ABA'B'}$ can be written as a $4\times 4$ array of blocks $\hat{A}_{ij,kl} = \bra[AB]{ij} \hat\varrho_{ABA'B'} \ket[AB]{kl}$:
\begin{equation}
\hat{\varrho}_{ABA'B'}=
\begin{pmatrix}
\hat{A}_{00,00} & \hat{A}_{00,01} & \hat{A}_{00,10} & \hat{A}_{00,11} \\
\hat{A}_{01,00} & \hat{A}_{01,01} & \hat{A}_{01,10} & \hat{A}_{01,11} \\
\hat{A}_{10,00} & \hat{A}_{10,01} & \hat{A}_{10,10} & \hat{A}_{10,11} \\
\hat{A}_{11,00} & \hat{A}_{11,01} & \hat{A}_{11,10} & \hat{A}_{11,11} \\
\end{pmatrix}
\label{initial-state}
\end{equation}
It can be transformed by LOCC (with respect to the partition $AA':BB'$) to the form:
\begin{widetext}
\begin{equation}
\hat{\tilde{\varrho}}=
\begin{pmatrix}
\frac{1}{2}(\hat{A}_{00,00} + \hat{A}_{11,11}) & 0  & 0  & \frac{1}{2}(\hat{A}_{00,11} + \hat{A}_{11,00}) \\
0 &  \frac{1}{2}(\hat{A}_{01,01} + \hat{A}_{10,10}) & \frac{1}{2}(\hat{A}_{01,10} + \hat{A}_{10,01})  & 0 \\
0 &  \frac{1}{2}(\hat{A}_{01,10} + \hat{A}_{10,01}) &  \frac{1}{2}(\hat{A}_{01,01}+ \hat{A}_{10,10})  & 0 \\
\frac{1}{2}(\hat{A}_{00,11} + \hat{A}_{11,00}) & 0 & 0 & \frac{1}{2}(\hat{A}_{00,00} + \hat{A}_{11,11})
\end{pmatrix}
\end{equation}

Which can be ,,unwisted'' to  Bell diagonal matrix which is directly related to privacy squeezed state (see \cite{smallkey,keyhuge}):
\begin{equation}
\hat{\sigma}_{AB}=
\begin{pmatrix}
\frac{1}{2}|| \hat{A}_{00,00} + \hat{A}_{11,11} || & 0  & 0  & \frac{1}{2} || \hat{A}_{00,11} + \hat{A}_{11,00}|| \\
0 &  \frac{1}{2} || \hat{A}_{01,01} + \hat{A}_{10,10} || & \frac{1}{2} || \hat{A}_{01,10} + \hat{A}_{10,01} || & 0 \\
0 &  \frac{1}{2} || \hat{A}_{01,10} + \hat{A}_{10,01} || &  \frac{1}{2} || \hat{A}_{01,01}+ \hat{A}_{10,10} ||  & 0 \\
\frac{1}{2} || \hat{A}_{00,11} + \hat{A}_{11,00}  || & 0 & 0 & \frac{1}{2}|| \hat{A}_{00,00} + \hat{A}_{11,11} ||
\end{pmatrix}
\label{Eq:sigmaABdef}
\end{equation}
\end{widetext}
where the norm $||\cdot||$ stands for the trace norm.

It will be convenient to write $\hat{\sigma}_{AB}$ in the form:
\begin{equation}
\hat{\sigma}_{AB}=
\begin{pmatrix}
\frac{1}{2}(p_1+p_2) & 0  & 0  & \frac{1}{2}(p_1-p_2) \\
0 & \frac{1}{2}(p_3+p_4) & \frac{1}{2}(p_3-p_4) & 0 \\
0 & \frac{1}{2}(p_3-p_4) & \frac{1}{2}(p_3+p_4) & 0 \\
\frac{1}{2}(p_1-p_2) & 0 & 0 & \frac{1}{2}(p_1+p_2) \\
\end{pmatrix}
\end{equation}
It is easy to see that $\hat{\sigma}_{AB}$ is diagonal in the Bell basis and the parameters $p_i$ are occupation probabilities
of the corresponding Bell states.

There is a useful bound on the secret key which is \cite{smallkey}
\begin{multline}
K_{D}(\hat\varrho_{ABA'B'})\geq I_{\text{cl}}(A:B)-S(\hat\sigma_{AB}) \\
=1- h(p_1 + p_2) - H(p_1,p_2,p_3,p_4)
\label{Eq:KD<=p1...p4}
\end{multline}
Here $I_\text{cl}(A:B)$ is the classical mutual information for the outcomes of von Neumann measurements carried out by Alice and Bob
in the secure basis $\ket[AB]{00}, \ket[AB]{01}, \ket[AB]{10}, \ket[AB]{11}$. As the joint probability distribution for these outcomes is
$\{ \frac{1}{2}(p_1+p_2), \frac{1}{2}(p_3+p_4), \frac{1}{2}(p_3+p_4), \frac{1}{2}(p_1+p_2) \} $, we have $I_{cl}(A:B)=1-h(p_1+p_2)$,
where $h(x)= - x \log_2  x - (1-x) \log_2 (1-x)$ denotes the binary entropy.

Note that the bound given in Eq.~(\ref{Eq:KD<=p1...p4}) is weaker than that on distillable entanglement in Eq.~(\ref{Eq:ED>=Bell}). This is because the state
$\hat{\sigma}_{AB}$ does not actually describe the physical system at any stage of the protocol, but is rather a formal construct
characterizing privacy properties of the state $\hat{\varrho}_{ABA'B'}$ of the complete system $ABA'B'$.

\section{Single privacy witness}
\label{sec:single}

The class of secrecy witnesses we shall consider here is defined formally as
\begin{align}
\hat{W}_{\text{key}} & = \bigl(\ket{11}_{AB}\bra{00} + \ket{00}_{AB}\bra{11} \bigr) \otimes \hat{U}_{A'B'} \nonumber \\
 & = (\hat\sigma^x_{A} \otimes \hat\sigma^x_{B} - \hat\sigma^y_{A} \otimes \hat\sigma^y_{B} )\otimes \hat{U}_{A'B'}
\label{witness}
\end{align}
where $\hat{U}_{A'B'}$ is any hermitian matrix satisfying $\hat{U}^\dagger \hat{U} \le \hat{\openone}$ acting on the shield subsystems $A'$ and $B'$.

We will use the expectation value of the secrecy witness
$\langle \hat{W}_{\text{key}} \rangle $ to approximate the value of $p_1-p_2$.
In fact we have:
\ben
w := | \langle \hat{W}_{\text{key}} \rangle |= \Tr[(\hat{A}_{00,11} + \hat{A}_{11,00})\hat{U}]  \nonumber \\
 \leq ||\hat{A}_{00,11} + \hat{A}_{11,00}|| =p_1 - p_2
\label{assumption}
\een
since for any matrix satisfying $\hat{U}^\dagger \hat{U} \le \hat{\openone}$ one has $\Tr(\hat{A}\hat{U})\leq ||\hat{A}||$. Indeed,
$\hat{U}^\dagger \hat{U} \le \hat{\openone}$ implies $||\hat{U}||_{\infty} \le1$ and consequently $\Tr\hat{A}\hat{U}\leq ||\hat{A}\hat{U}||\leq ||\hat{A}||\cdot||\hat{U}||_{\infty}$. Sometimes the parameter $w$ may give exactly the value of $p_1 - p_2$. We will give an instance of that at the end of this section. Let us also stress  that the witness itself (\ref{witness}) is an observable which must be measured on the original state (\ref{initial-state}).

Quantitative estimation of the distillable key will be based on the following inequality
\begin{equation}
H(p_1,p_2,p_3,p_4) \leq
H \left( p_1,p_2,{\textstyle\frac{1}{2}}(1-p_1-p_2),{\textstyle\frac{1}{2}} (1-p_1-p_2) \right)
\end{equation}
applied to Eq.~(\ref{Eq:KD<=p1...p4}), which yields:
\begin{multline}
K_{D}
   \ge 1-h(p_1+p_2) \\ - H \left( p_1,p_2,{\textstyle\frac{1}{2}}(1-p_1-p_2),{\textstyle\frac{1}{2}} (1-p_1-p_2) \right).
\label{bound1}
\end{multline}
As we are interested in the most pessimistic scenario, we need to minimize the right hand side over pairs $(p_1, p_2)$ that satisfy all the constraints. This gives the {\it central formula}:
\begin{multline}
K_{D} \geq 1-\sup_{\substack{p_1-p_2 \geq w \\ p_1, p_2 \ge 0, p_1 +p_2 \le 1}} \left[ h(p_1+p_2) \vphantom{\textstyle\frac{1}{2}} \right. \\
+ H \left. \left( p_1,p_2,{\textstyle\frac{1}{2}}(1-p_1-p_2),{\textstyle\frac{1}{2}} (1-p_1-p_2) \right) \right]
\label{CentralBound}
\end{multline}

We found numerically the boundary value $w^*$ at which the above bound becomes strictly positive,
i.e.\ the witness condition $w=| \langle \hat{W}_{\text{key}} \rangle|  > w^\ast$, to be equal to $w^\ast \approx 0.907$.

We can simplify the optimization in Eq.~(\ref{CentralBound}) by introducing a new pair of variables $p_+ = p_1 + p_2$ and $\xi_+ = p_1/(p_1+p_2)$. A straightforward calculation shows that the bound for
the key expressed in the new variables takes the form:
\begin{equation}
K_D \geq \inf_{\substack{w \le p_+ \le 1 \\ (w + p_+)/2p_+ \le \xi_+ \le 1}} \bigl( p_+ - 2 h(p_+) - p_+ h(\xi_+) \bigr).
\label{Eq:KD>=p+xi+}
\end{equation}
Because the lower limit for $\xi_+$ is greater or equal to $1/2$, optimization over $\xi_+$ for a fixed value of $p_+$ yields $h(\xi_+) \le h\bigl((w + p_+)/2p_+\bigr)$. Consequently, the minimization in Eq.~(\ref{Eq:KD>=p+xi+}) needs to be carried out over a single parameter $p_+$:
\begin{equation}
K_D \geq \inf_{w \le p_+ \le 1 } \bigl[ p_+ - 2 h(p_+) - p_+ h \bigl((w + p_+)/2p_+\bigr) \bigr].
\end{equation}
The absolute minimum of this expression is analyzed in Appendix.
However we can simplify the bound in two ways leading to weaker but more compact formulas:

{\it Weaker bound 1:\/}
Suppose that $w \geq (1-w)/3$ which is equivalent to $w\geq 1/4$.
Then we have the following estimate:
\begin{equation}
 K_{D} \geq 1  - h(w) - H \left( w, {\textstyle\frac{1}{3}}(1-w),
 {\textstyle\frac{1}{3}}(1-w), {\textstyle\frac{1}{3}}(1-w) \right)
\label{weaker-bound-1}
\end{equation}
In the last inequality we have used the fact that both $p_1+p_2 \ge w$ and $p_1\geq w$.

{\it Weaker bound 2:\/}
There is a possibility of having another bound with help
of subadditivity of the entropy $H[p_1,p_2,p_3,p_4] \leq h(p_1+p_2) + h(p_1+p_3)$.
\begin{eqnarray}
 K_{D} \geq  1  - 2 h(w) - h\left({\textstyle\frac{1}{2}}(1+w)\right)
 \label{weaker-bound-2}
\end{eqnarray}

\begin{figure}[h!]
  \centering
      \includegraphics[width=6cm]{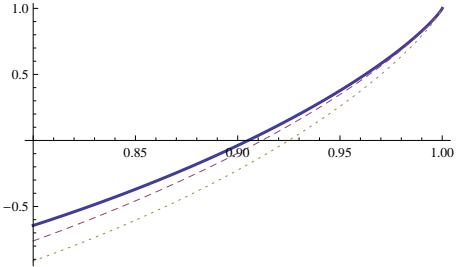}
  \caption{Graphs of lower bounds on the distillable key as a function of $w= | \langle \hat{W}_{\text{priv}} \rangle|$
  derived in Eqs.~(\protect\ref{CentralBound}) (solid line), (\protect\ref{weaker-bound-1}) (dashed line),
  and (\protect\ref{weaker-bound-2}) (dotted line).}
  \label{Fig:SingleWitness}
\end{figure}

For a graphic comparison of the derived formulas see Fig.~\ref{Fig:SingleWitness}.

The above considerations are based on the ccq scenario, when Alice and Bob measure their qubits in the secure key basis $|0\rangle$, $|1\rangle$.
However if the shield part is not affected by local unitary the optimal $\hat{U}$ remains unchanged. For instance, in the case of the four-qubit state whose approximate version was realized experimentally in Ref.~\cite{pbits-exp}, a two-qubit swap operator $\hat{V}_{A'B'} = \hat{\openone}_{A'B'} - 2 \proj[A'B']{\Psi_-}$ used in the privacy witness would give {\it exactly} the value $| \langle \hat{W}_{\text{priv}}\rangle |= p_1-p_2$.
Note also  that if $\hat{U}$ were not hermitian, we could decompose $\hat{U} = \hat{U}_R + i \hat{U}_I$ and measure
 two observables
$\hat{W}_{\text{key}}^R= (|11\>_{AB}\<00|+|00\>_{AB}\<11|)\ot \hat{U}_R$ and $\hat{W}_{\text{key}}^I= (|11\>_{AB}\<00|+|00\>_{AB}\<11|)\ot \hat{U}_I$.

\section{Two-observable privacy estimation}
\label{ref:double}

In this section we will show how to characterize the privacy properties of a noisy entangled state from expectation values of two observables. The first one, given by $\hat{\sigma}^z_{A} \otimes  \hat{\sigma}^z_{B} \otimes \hat{\openone}_{A'B'}$, characterizes correlations between measurements performed on the subsystems $A$ and $B$ in the key basis. The security of the key will be inferred using the expectation value of an observable $\hat{\sigma}^x_{A} \otimes  \hat{\sigma}^x_{B} \otimes \hat{U}_{A'B'}$.

In the further discussion it will be convenient to use the following parameterization of $\hat{\sigma}_{AB}$:
\begin{equation}
\hat{\sigma}_{AB} =
\begin{pmatrix}
\frac{1}{2} p_+ & 0 & 0 & p_+(\xi_+ -\frac{1}{2}) \\
0 & \frac{1}{2} p_- & p_- (\xi_- - \frac{1}{2}) & 0 \\
0 & p_- (\xi_- - \frac{1}{2}) & \frac{1}{2} p_-  & 0 \\
 p_+(\xi_+ -\frac{1}{2}) & 0 & 0 & \frac{1}{2}p_+
\end{pmatrix}.
\end{equation}
The nonnegative parameters $p_+ = p_1 + p_2$ and $p_- = 1-p_+ = p_3 + p_4$ can be interpreted as occupation probabilities of the correlated and the anticorrelated subspaces, spanned respectively by pairs or vectors $\ket[AB]{00}$, $\ket[AB]{11}$ and $\ket[AB]{01}, \ket[AB]{10}$. The other two parameters, given explicitly by $\xi_+ = p_1/p_+$ and $\xi_- = p_3/p_-$, characterize the off-diagonal elements of $\hat{\sigma}_{AB}$ respectively in the correlated and the anticorrelated sectors and therefore contain information about the privacy properties. Because the off-diagonal elements of $\hat{\sigma}_{AB}$ are nonnegative due to the definition given in Eq.~(\ref{Eq:sigmaABdef}) and must ensure positive definiteness of $\hat{\sigma}_{AB}$, the parameters $\xi_-,\xi_+$ satisfy the inequality:
\begin{equation}
\frac{1}{2} \le \xi_+, \xi_- \le 1,
\label{Eq:squareofxis}
\end{equation}
i.e.\ the relevant region for pairs $(\xi_-,\xi_+)$ has the shape of a square.

In the new parameterization, the lower bound on the key takes the following form:
\begin{equation}
\label{Eq:KDbound}
K_D \ge 1 - 2 h(p_+) - p_+ h(\xi_+) - p_- h(\xi_-).
\end{equation}
Because the binary entropies $h(\xi_+)$ and $h(\xi_-)$ are nonnegative, a necessary condition for this bound to be nontrivial is $h(p_+) < \frac{1}{2}$, otherwise the right hand side is not positive. This means that $p_+$ must satisfy either $0 \le p_+ < 1-p^\ast$ or $p^\ast < p_+ \le 1$, where $p^\ast \approx 0.89$ is the bigger of two solutions of a transcedental equation $h( p^\ast) = \frac{1}{2}$ on the interval $ 0 \le p^\ast \le 1$.
We will restrict our further discussion only to the case $p^\ast < p_+ \le 1$, as the analysis of the second case $0 \le p_+ < 1-p^\ast$ is completely analogous.

Let us now discuss how the parameters of $\hat{\sigma}_{AB}$ are related to measured observables.
The parameters $p_\pm$ can be evaluated directly from the measured observables as
$p_\pm = \frac{1}{2} \left(1 \pm \langle \hat{\sigma}^z_{A} \otimes  \hat{\sigma}^z_{B} \otimes \hat{\openone}_{A'B'} \rangle \right)$.
Following discussion after Eq.~(\ref{Eq:KDbound}), we will be interested in the regime when $p_+ > p^\ast$.
Considering the other regime when $p_+ < 1- p^\ast$ effectively boils down to swapping the roles of the correlated and the anticorrelated subspaces. These two possibilities can be analyzed jointly by defining
\begin{equation}
w_z := \left| \langle \hat{\sigma}^z_{A} \otimes  \hat{\sigma}^z_{B} \otimes \hat{\openone}_{A'B'} \rangle \right|
\end{equation}
and using in further discussion
\begin{equation}
p_\pm = \frac{1}{2} (1 \pm w_z).
\end{equation}
The condition $p_+ > p^\ast$ can be equivalently written as
\begin{equation}
w_z  > 2p^\ast -1 \approx 0.78
\end{equation}
which defines the minimum value of $w_z$ above which the bound on the key can become nontrivial.

The second quantity we will use in our analysis will be
\begin{equation}
w_x  := \bigl| \langle \hat{\sigma}^x_{A} \otimes  \hat{\sigma}^x_{B} \otimes \hat{U}_{A'B'} \rangle \bigr|.
\end{equation}
It allows us to bound
the parameters $\xi_-$ and $\xi_+$ according to the following inequality which will be at the heart of our reasoning:
\begin{align}
w_x & = \bigl| \Tr [\hat{U}(\hat{A}_{00,11}+ \hat{A}_{01,10} + \hat{A}_{10,01} + \hat{A}_{11,00})] \bigr|
\nonumber \\
&\le ||\hat{A}_{00,11} + \hat{A}_{11,00} || + || \hat{A}_{01,10} + \hat{A}_{10,01} || \nonumber \\
& = p_+(2\xi_+ -1) + p_-(2\xi_- -1).
\label{Eq:Cxp1p2p3p4}
\end{align}
For fixed $p_\pm$, this inequality determines the allowed region of $(\xi_-,\xi_+)$ in the square defined by Eq.~(\ref{Eq:squareofxis}) as
\begin{equation}
\label{Eq:p+xi+p-xi-}
p_- \xi_- + p_+\xi_+  \ge {\textstyle \frac{1}{2}} ( 1 + w_x ).
\end{equation}
When evaluating the lower bound on the key rate $K_D$ according to Eq.~(\ref{Eq:KDbound}) we are interested in the worst case scenario that is consistent with the measurement results. Therefore our task is to minimize the right hand side of Eq.~(\ref{Eq:KDbound}) under constraints given by Eqs.~(\ref{Eq:squareofxis}) and (\ref{Eq:p+xi+p-xi-}). This is equivalent to maximizing under the same constraints a concave function
\begin{equation}
\label{Eq:fxi-xi+def}
f(\xi_-, \xi_+) =  p_- h(\xi_-) + p_+ h(\xi_+) .
\end{equation}
The lower bound on the key can be written as
\begin{equation}
\label{Eq:KD=fmax}
K_D \ge 1 - 2 h(p_+) - f^{\text{max}},
\end{equation}
where $f^{\text{max}}$ is the maximum of $f(\xi_-, \xi_+)$ over the allowed region of parameters. It is useful to note that
because $f(\xi_-, \xi_+)$ is a convex linear combination of binary entropies $h(\xi_-)$ and $h(\xi_+)$, within the square given by Eq.~(\ref{Eq:squareofxis}) decreasing either of the arguments $\xi_-$ or $\xi_+$ will always increase the value of $f(\xi_-, \xi_+)$. This in turn implies that $f^{\text{max}}$ is reached on the line
\begin{equation}
\label{Eq:linexi-xi+}
p_- \xi_- + p_+\xi_+  = {\textstyle \frac{1}{2}} ( 1 + w_x ).
\end{equation}

\begin{figure}
\includegraphics[width=3.375in]{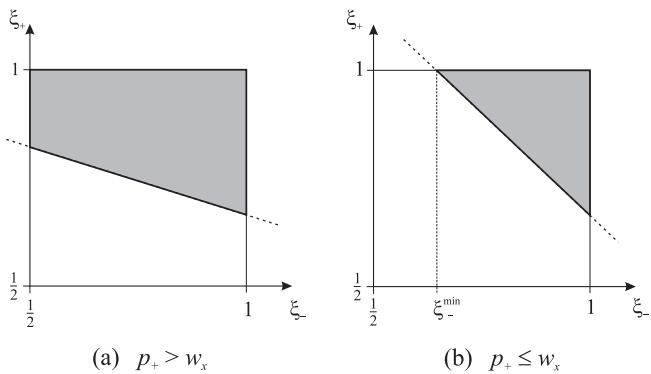}
\caption{The permitted region of the parameters $(\xi_-,\xi_+)$ used to maximize the function $f(\xi_-,\xi_+)$
defined in Eq.~(\protect\ref{Eq:fxi-xi+def}).}
\label{Fig:xisquare}
\end{figure}

\begin{figure}
\includegraphics[width=3in]{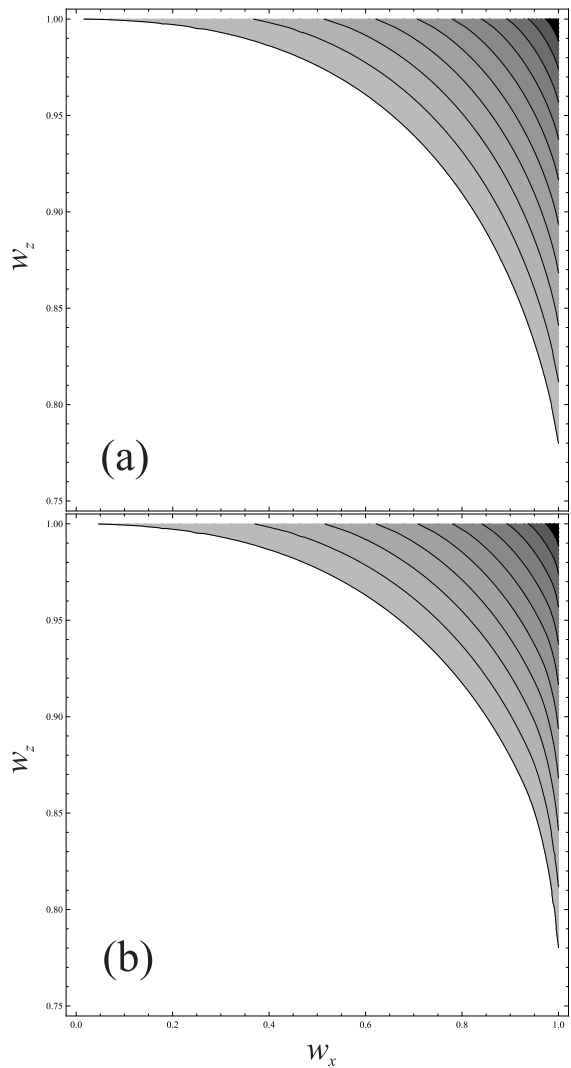}
\caption{The lower bound on the key as a function of $w_x$ and $w_z$ obtained from (a) full optimization over the free parameters
specified in Eq.~(\protect\ref{Eq:KDfullbound}) and (b) a weaker estimate according to Eq.~(\protect\ref{Eq:KDWeakerbound}).
The bound is positive in the shaded area, with solid contours drawn at steps $0.1$ starting from $0$.}
\label{Fig:wxqz}
\end{figure}

To proceed with the maximization, let us start from an observation
that if $\xi_+ = \frac{1}{2}$, i.e.\ the correlated sector of the density matrix has zero off-diagonal elements, no positive key rate can be guaranteed by Eq.~(\ref{Eq:KDbound}). This follows from a straightforward fact that the expression $1 - 2 h(p_+) - p_+$ is nonpositive for $p^\ast < p_+ \le 1$. Therefore no point with $\xi_+ = \frac{1}{2}$ should satisfy Eq.~(\ref{Eq:p+xi+p-xi-}). Because the slope of the line (\ref{Eq:linexi-xi+}) is negative, it is sufficient to require that the point $(\xi_-=1, \xi_+ = \frac{1}{2})$ is outside the allowed region. This is equivalent to the inequality:
\begin{equation}
w_x + w_z > 1.
\end{equation}
Further analysis depends on whether the point $(\xi_- = \frac{1}{2}, \xi_+ = 1)$ is located within the allowed region of parameters. It is easy to verify that this is determined by the relation between $p_+$ and $w_x$. If $p_+ > w_x$, this point satisfies Eq.~(\ref{Eq:p+xi+p-xi-}) and the allowed region of parameters has the shape of a trapezoid, as shown in Fig.~\ref{Fig:xisquare}(a). Consequently, all values $\frac{1}{2} \le \xi_- \le 1$ are allowed.
On the other hand, when $p_+ \le w_x$, the allowed region is a triangle, as depicted in Fig.~\ref{Fig:xisquare}(b).
The minimum allowed value of $\xi_-$ is then $( w_x - w_z )/(1-w_z)$. We can combine these two cases by defining
\begin{equation}
\label{Eq:xi-min}
\xi_-^{\text{min}} = \max \left\{ \frac{1}{2}, \frac{w_x - w_z}{1 - w_z}\right\}
\end{equation}
The maximization of $f(\xi_-,\xi_+)$ over the line defined in Eq.~(\ref{Eq:linexi-xi+}) can now be written as a supremum over a single parameter:
\begin{equation}
f^{\text{max}} = \sup_{ \xi_-^{\text{min}} \le \xi \le 1} f \left( \xi, \frac{1+ w_x}{1 + w_z} - \xi \frac{1 - w_z}{1 + w_z} \right)
\end{equation}
which inserted into Eq.~(\ref{Eq:KD=fmax}) yields the final form of the bound:
\begin{multline}
K_D \ge 1 - 2 h(p_+) \\ - \sup _{ \xi_-^{\text{min}} \le \xi \le 1} \left[
(1-p_+) h(\xi) + p_+ h\left(\frac{1+ w_x}{1 + w_z} - \xi \frac{1 - w_z}{1 + w_z} \right)
\right]
\label{Eq:KDfullbound}
\end{multline}
where $p_+ = (1+w_z)/2$. The results of a numerical evaluation of the supremum are shown in Fig.~\ref{Fig:wxqz}(a).

It is also possible to derive a slightly weaker bound that requires no numerical maximization, which we will denote as $\tilde{f}^{\text{max}}$. Because the function $f(\xi_-, \xi_+)$ is monotonic in each one of its two arguments, as discussed after Eq.~(\ref{Eq:KD=fmax}), we have
\begin{equation}
f^{\text{max}} \le \tilde{f}^{\text{max}} = f(\xi_-^{\text{min}}, \xi_+^{\text{min}}).
\end{equation}
where $\xi_-^{\text{min}}$ and $\xi_+^{\text{min}}$ are the smallest possible values of $\xi_-$ and $\xi_+$ within the allowed region defined by Eqs.~(\ref{Eq:squareofxis}) and (\ref{Eq:p+xi+p-xi-}). The value $\xi_-^{\text{min}}$ has been given explicitly in Eq.~(\ref{Eq:xi-min}), while it is easy to verify that
\begin{equation}
\xi_+^{\text{min}} = \frac{ w_x + w_z }{1 + w_z}.
\end{equation}
The simplified bound therefore takes the following explicit form:
\begin{equation}
K_D \ge 1 - 2 h(p_+) - (1-p_+) h\left(\xi_-^{\text{min}}\right) - p_+ h \left( \frac{ w_x + w_z }{1 + w_z} \right)
\label{Eq:KDWeakerbound}
\end{equation}
which is shown in Fig.~\ref{Fig:wxqz}(b).

\begin{figure}
\includegraphics[width=3in]{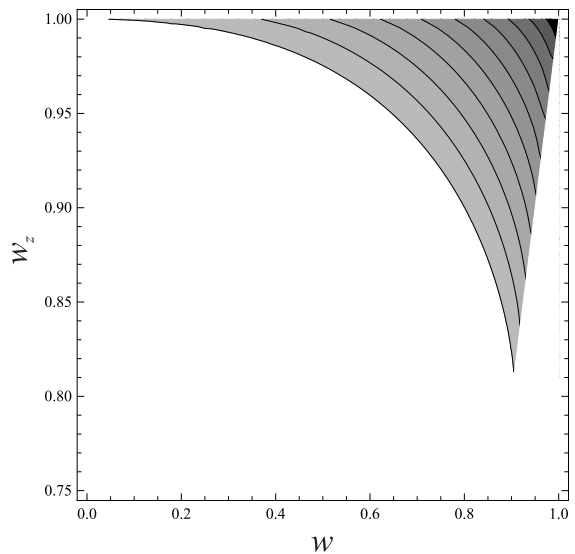}
\caption{The lower bound on the key given in Eq.~(\protect\ref{Eq:KDwwz}) obtained from observables $w$ and $w_z$. The physical region is
restricted to points $(w,w_z)$ satisfying the condition $(1+w_z)/2 \ge w $. The coding of the bound value is the same as in Fig.~\protect\ref{Fig:wxqz}.}
\label{Fig:KDwwz}
\end{figure}

Finally, let us note that the observable $w$ defined in Eq.~(\ref{assumption}) can be combined with $w_z$ to yield a stronger bound on the distillable key. To derive this bound, let us return to Eq.~(\ref{Eq:KDbound}) and estimate the last two terms on the right hand side. The inequality shown in Eq.~(\ref{assumption}) rewritten in the new parametization provides a lower bound on $\xi_+$:
\begin{equation}
\xi_+ \ge \frac{1}{2} + \frac{w}{2 p_+}.
\end{equation}
Because the right hand side is greater or equal $1/2$, we have $h(\xi_+) \le h ( 1/2 +w/2 p_+ )$. Further, we obviously have $h(\xi_-) \le 1$. This yields:
\begin{equation}
K_D \ge p_+ \left[ 1- h\left(  1/2 +w/2 p_+ \right) \right] - 2 h \left( p_+ \right).
\label{Eq:KDwwz}
\end{equation}
Let us note that physical values of $w$ and $w_z$ must satisfy the condition $(1+w_z)/2 \ge w $, otherwise we would have $p_1 + p_2 = p_+ = (1+w_z)/2  < w \le p_1 -p_2$ and consequently $p_2 < 0$. In Fig.~\ref{Fig:KDwwz} we depict the bound (\ref{Eq:KDwwz}) in the physical region of $w$ and $w_z$.

\section{Discussion and conclusions}
\label{Sec:Conclusions}

In this paper we have introduced a privacy witness, i.e.\ an observable, whose mean value allows one to estimate nontrivially from below
the amount of secret key, even in the case when the resulting state exhibits the separation of the secret key and the distillable entanglement.
In fact this separation may be quite drastic while the witness is working perfectly.
To see this let us take  $A'B'$ being $d \otimes d $-level systems and consider a p-bit state $\varrho_{ABA'B'}$
in the form \cite{pptkey}
\begin{align}
\hat\varrho_{ABA'B'} & = \frac{1}{2d^2} \left[ \left( \proj[AB]{00} + \proj[AB]{11} \right) \otimes \hat{\openone}_{A'B'} \right. \nonumber \\
& + \left. \left( \ket[AB]{00}\bra{11} + \ket[AB]{11}\bra{00} \right) \otimes \hat{V}_{A'B'} \right]
\end{align}
where $\hat{\openone}_{A'B'}$ and $\hat{V}_{A'B'}$ stand respectively for bipartite identity and swap operator on the subsystems $A'B'$.
In the limit of large $d$ the distillable entanglement is bounded by vanishing log-negativity $E_{D}\leq \log_2 (1+\frac{1}{d}) \rightarrow 0$,  while
the privacy witness $W=(\hat\sigma^x_A \otimes \hat\sigma^x_B + \hat\sigma^y_A \otimes  \hat\sigma^y_B) \otimes \hat{V}_{A'B'}$
gives us the value of lower bound $K_{D}\geq 1$ since  $w=Tr( \hat{W} \hat{\varrho}_{ABA'B'})=1$
and then just either of the weaker bounds given in Eqs.~(\ref{weaker-bound-1}) or (\ref{weaker-bound-2})
does the job.
Note in particular that since the key part $AB$ is a two qubit part the above estimate gives the maximum possible value of the secret key
$K_{D}=\log_2 2=1$ despite the fact that the distillable entanglement of the state is almost zero!

In general the complexity of measuring the privacy witness is related to the Hilbert-Schmidt decomposition of the
hermitian operator $\hat{U}$ used to construct the witness \cite{GuehneLocalMeasurementWitness}. In the case of the four qubit state
that was studied in the experiment reported in Ref.~\cite{pbits-exp}
the operator in question is the swap operator which is composed of three terms involving products of Pauli matrices:
\begin{equation}
\hat{V}_{A'B'} = \frac{1}{2} ( \hat{ \openone}_{A'B'} + \hat{\sigma}_{A'}^{x} \otimes \hat{\sigma}_{B'}^{x}
+  \hat{\sigma}_{A'}^{y} \otimes \hat{\sigma}_{B'}^{y} +  \hat{\sigma}_{A'}^{z} \otimes \hat{\sigma}_{B'}^{z}).
\end{equation}
Taking into account the necessary measurements on the key part this gives in total
$2 \times 3 = 6$  observables to be measured, each formed by a product of four Pauli matrices.
This is dramatically less then the full tomography which requires $81$ products of four Pauli matrices.
Note that in some cases, like the p-bit state discussed above, such an apparently poor measurement
has no problem in reporting the key value that lies above the distillable entanglement,
which is bounded for our example by the log negativity value $E_{D}(\varrho) \leq \log_2 (1+\frac{1}{2}) \approx 0.585$.

The above approach may be extended to higher dimensions and other twirling techniques may be applied.
It may be especially useful when the experimentalist has a good guess about the expected p-bit he state
in the laboratory then he or she may estimate the high key contents almost perfectly even if there is virtually
no distillable entanglement in the system.
Finally let us note that the very difficult problem is to find the nonlinear entanglement witness that
would capture collective behavior revealing the key in all the cases when any single-copy entropic fuction
based on a one-way protocol fails. It seems that for this one needs quantum secrecy distillation protocols of new generation.

We believe that the present approach will lead to general and systematic
development of experimentally friendly methods for detection and estimation
of task-dedicated quantum entanglement and other resources.

\begin{acknowledgments}

This work was supported by 7th Framework Programme Future and Emerging Technologies projects CORNER (no.\ 213681) and Q-ESSENCE (no.\ 248095).
K.H. is supported by grant BW no. 538-5300-0637-1,  P.H. is supported by NN 202-261-938. We acknowledge useful discusions with R. Demkowicz-Dobrza\'{n}ski.

\end{acknowledgments}

\appendix

\section{Single witness bound}

\begin{figure}[h!]
\includegraphics[width=3in]{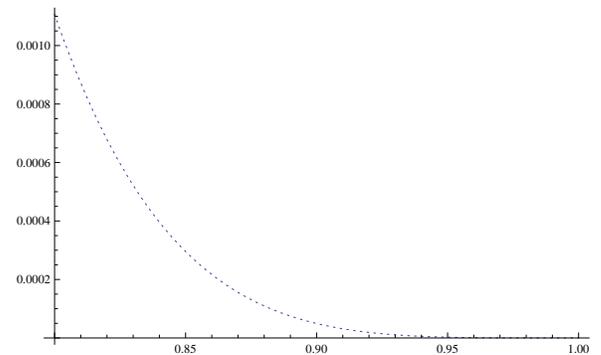}
\caption{The difference between the bounds from Eqs.~(\protect\ref{Eq:KD>=p+xi+}) and (\protect\ref{Eq:KDapprox}) as a function of the parameter $w$. }
\label{Fig:difference}
\end{figure}

Let us denote
\begin{equation}
\kappa (p_+) =   p_+ - 2 h(p_+) - p_+ h \bigl((p_+ +w)/2p_+\bigr).
\end{equation}
The derivative with respect to $p_+$ reads:
\begin{equation}
\frac{\text{d}\kappa}{\text{d}p_+} = \frac{1}{2} \log_2 \frac{p_+^2 (p_+^2 - w^2)}{(1-p_+)^4}.
\end{equation}
It is easy to see that on the interval $w \le p_+ \le 1$ the argument of the logarithm function runs from
$0$ to $+\infty$. Therefore $\kappa (p_+)$ reaches its minimum at a root of a polynomial equation
$p_+^2 (p_+^2 - w^2) = (1-p_+)^4$. This is a cubic equation which can be solved exactly, but a simplified formula can be found
by substituting $p_+ = w+\delta$ and assuming that $\delta \ll 1-w$ which is motivated by numerical analysis. This yields
$\delta \approx (1-w)^4/2w^3$ and consequently the bound on the key in the approximate form
\begin{equation}
K_D \ge \kappa \left( w + \frac{(1-w)^4}{2w^3} \right).
\label{Eq:KDapprox}
\end{equation}
This approximate expression turns out to reproduce the original bound quite tightly, as evidenced in Fig.~\ref{Fig:difference} depicting the difference between Eq.~(\ref{Eq:KDapprox}) and the bound given in Eq.~(\ref{Eq:KD>=p+xi+}).

\bibliographystyle{apsrev}

\end{document}